\documentclass[twocolumn,showpacs,groupedaddress]{revtex4}
\usepackage{graphicx}
\usepackage{epsfig}
\usepackage{subfigure}
\usepackage{amsmath}

\begin{document}

\newcommand{\ket}[1]{|#1\rangle}
\newcommand{\bra}[1]{\langle#1|}

\title{
Driving Quantum System into Decoherence-free Subspaces  by Lyapunov
Control}
\author{ X. X. Yi$^{1,2}$, X. L. Huang$^1$, Chunfeng  Wu$^2$, and C. H. Oh$^2$}

\affiliation{$^1$School of Physics and Optoelectronic Technology,
Dalian University of Technology, Dalian 116024, China \\
$^2$Centre for Quantum Technologies and Department of Physics,
National University of Singapore, 3 Science Drive 2, Singapore
117543, Singapore}

\begin{abstract}
We present a scheme  to drive a finite-dimensional quantum system
into the decoherence-free subspaces(DFS) by Lyapunov control.
Control fields are established by Lyapunov function. This proposal
works well for both closed and open quantum systems, with replacing
the DFS by desired subspaces for closed systems. An example which
consists of a four-level system with three degenerate states driven
by three lasers is presented to gain further insight of the scheme,
numerical simulations for the dynamics of the system  are performed
and the results are good.
\end{abstract}

 \maketitle

\section{introduction}

Quantum computing  and quantum communication  have attracted a lot
of attention due to their  promising applications such as the
speedup of classical computations and secure  key
distributions~\cite{chuang95}. Although the physical implementation
of basic quantum information processors has been reported
recently~\cite{monroe02}, the realization of powerful and useable
devices is still a challenging and as yet unresolved task. A major
difficulty arises from the coupling  of a quantum system to its
environment that leads to decoherence. One promising solution to
this problem is provided by the concept of decoherence-free
subspaces (DFS)~\cite{zanardi97}. The decoherence-free subspaces
have been defined as collections of states that  undergo unitary
evolution in the presence of decoherence. Experimental realizations
of DFS have been achieved with photons~\cite{kwiat00} and in nuclear
spin systems~\cite{viola01}. A decoherence-free quantum memory for
one qubit has been realized experimentally with two trapped
ions~\cite{kielpinski01,langer05}. As the DFS is a promising
candidate to solve the problem of decoherence, it is natural to ask
how can we drive a quantum system into the DFS by quantum control?

Several approaches to controlling
\cite{rabitz00,lloyd00,ramakrishna95,schirmer01,
wiseman94,mancini98,belavkin99,doherty99,carvalho07,yi08}
 a quantum system have been proposed in the past decade, which can
be divided into coherent (unitary) and incoherent (non-unitary)
control, according to how the controls enter the dynamics. In the
coherent control scheme, the controls enter the dynamics through the
system Hamiltonian. It affects the time evolution of the system, but
not its spectrum. In the incoherent(but linear) control scheme
\cite{romano06,xue06,nie08}, an auxiliary system, called probe, is
introduced to manipulate the target system through their mutual
interactions. This incoherent control scheme is of relevance
whenever the system dynamics cannot be directly accessed, and it
provides a non-unitary  evolution  capable for transferring all
initial states (pure or mixed) into an arbitrary (pure or mixed)
target state.  While the above control strategies render  the
quantum system a linear dynamics, feedback control can lead the
dynamics
 to both nonlinear and
stochastic\cite{habib06}. In certain special cases, the dynamics can
be mapped into a linear classical system driven by Gaussian noise,
however, this method is not applicable to most quantum systems under
feedback control, and the experience from nonlinear control theory
in classical system tells us that it is unlikely to manipulate
optimally  a quantum system into a specific target state by  such a
control. Nevertheless, it may be possible to manipulate the quantum
system into a collection of states, for example the decoherence-free
subspaces.

In this paper, we  explore the Lyapunov control to manipulate an
open quantum system through driving it into the DFS. The Lyapunov
control has been proven to be a sufficient  simple control to be
analyzed rigorously, in particular,  this control can be shown to be
highly effective for systems that satisfy certain sufficient
conditions, which roughly speaking are equivalent to the
controllability of the linearized system. In Lyapunov control,
Lyapunov
 functions which were
originally used in feedback control to analyze the stability of the
control system,  have formed the basis for new control design. By
properly choosing  the Lyapunov function, we illustrate by an
example that  quantum systems can be controlled into the DFS.

This paper is organized as follows. In Sec. \ref{gf}, we present a
general analysis of Lyapunov control for open quantum systems,
Lyapunov functions and control fields  are given and discussed. To
illustrate the general formulism, we exemplify a four-level system
with 2-dimensional DFS in Sec.\ref{exa}, showing that  the system
can be driven in the DFS  by  Lyapunov  control. Finally, we
conclude our results in Sec. \ref{con}.

\section{general formulism}\label{gf}
A controlled quantum system can be modeled in different ways, either
as a closed system evolving unitarily governed  by a Hamiltonian, or
as an open system coupling to its environment. In this paper, we
restrict our discussion to a $N$-dimensional open quantum system,
and consider its dynamics  as Markovian and therefore the dynamics
obeys the Markovian master equation ($\hbar=1,$ throughout this
paper),
\begin{eqnarray}
\dot{\rho}&=&-i[H,\rho]+{\mathcal L}(\rho),\nonumber\\
{\mathcal L}(\rho)&=&\frac 1 2\sum_{m=1}^M \lambda_m([J_m,\rho
J_m^{\dagger}]+[J_m\rho,J_m^{\dagger}]),\nonumber\\
H&=&H_0+\sum_{n=1}^F f_n(t)H_n, \label{mef}
\end{eqnarray}
where $\lambda_m (m=1,2,...,M)$ are positive and time-independent
parameters, which characterize the decoherence. $J_m (m=1,2,...,M)$
are jump operators. $H_0$ is a free Hamiltonian and $H_n
(n=1,2,...,F)$ are control Hamiltonian, while $f_n(t) (n=1,2,...,F)$
are control fields. Equation (\ref{mef}) is of Lindblad form, this
means that the solution to Eq. (\ref{mef}) has all the required
properties of a physical density matrix at all times.

By its definition, DFS is composed  of states that undergo unitary
evolution. Considering the fact that there are many ways for a
quantum system to evolve unitarily, we  focus in this paper on the
DFS for which the dissipative part ${\mathcal {L}(\rho)}$ of the
master equation is zero, leading to the following conditions for
DFS\cite{karasik08}. A space spanned by $\mathcal{H}_{DFS}=\{
|\psi_1\rangle, |\psi_2\rangle, ..., |\psi_D\rangle \}$ is a
decoherence-free subspace for all time $t$ if and only if (1)
$\mathcal{H}_{DFS}$ is invariant under $H_0$; (2)
$J_m|\psi_n\rangle=c_m|\psi_n\rangle$ and (3)
$\Gamma|\psi_n\rangle=g|\psi_n\rangle$ for all $n=1,2,...,D$ and
$m=1,2,...,M$ with $g=\sum_{l=1}^M\lambda_l|c_l|^2,$ and
$\Gamma=\sum_{m=1}^M\lambda_mJ_m^{\dagger}J_m.$ With these
notations, the goal of this paper can be formulated as follows. We
wish to apply a specified set of control fields $\{ f_i(t),
n=1,2,...,F\}$ in Eq. (\ref{mef}) such that $\rho(t)$ evolves into
the DFS and stays there forever. In contrast to the conventional
control problem\cite{wangx09}, we here do not specify the target
state, instead the resulting state being in DFS is desired. Since
the free Hamiltonian $H_0$ cannot be turned off in general, we
assume that the resulting state $\rho_D=\rho_D(t) \in
\mathcal{H}_{DFS}$ is time-dependent and satisfies,
\begin{equation}
\dot{\rho}_D(t)=-i[H_0, \rho_D(t)].
\end{equation}
The control fields $\{f_n(t)\}$ can be established by Lyapunov
function. Define a function $V(\rho_D, \rho)$ as
\begin{equation}
V(\rho_D,\rho)=\mbox{Tr}(\rho_D^2)-\mbox{Tr}(\rho\rho_D),
\end{equation}
we find $V\geq 0$ and
\begin{equation}
\dot{V}=-\sum_n^F f_n(t)
\mbox{Tr}\{\rho_D[-iH_n,\rho]\}-\mbox{Tr}[\rho_D\mathcal{L}(\rho)].
\end{equation}
For $V$ to be a Lyapunov function, it requires  $\dot{V}\leq 0$ and
$V\geq 0.$ If we choose a $n_0$ such that $f_{n_0}(t)
\mbox{Tr}\{\rho_D[-iH_{n_0},\rho]\}+\mbox{Tr}[\rho_D\mathcal{L}(\rho)]=0$\cite{note1},
and $f_n(t)=\mbox{Tr}\{\rho_D[-iH_n,\rho]\}$ for $n\neq n_0$, then
$\dot{V}\leq 0.$  With these choices, $V$ is a Lyapunov function.
Therefore,  the evolution of the open system with Lyapunov feedback
control  described by the following nonlinear equations
\begin{eqnarray}
\dot{\rho}(t)&=&-i[H_0+\sum_{n} f_n(t)H_n,
\rho(t)]+\mathcal{L}(\rho),\nonumber\\
f_n(t)&=&\mbox{Tr}\{[-iH_n,\rho]\rho_D\}, \mbox{ for } n\neq
n_0,\nonumber\\
f_{n_0}(t)&=&-\frac{\mbox{Tr}[\rho_D\mathcal L
(\rho)]}{\mbox{Tr}\{\rho_D[\rho, iH_{n_0}]\}},\mbox{and} \nonumber\\
\dot{\rho}_D(t)&=&-i[H_0,\rho_D(t)]
\label{nle1}
\end{eqnarray}
is stable in Lyapunov sense at least.  Notice that the choice of
Lyapunov function is not unique, we may define the other  Lyapunov
function via the bases of DFS as,
\begin{equation}
V_b(\{|\psi_j\rangle\},\rho)=\frac 1
D(1-\sum_{j=1}^D\langle\psi_j^t|\rho|\psi_j^t\rangle),
\end{equation}
where $|\psi_j^t\rangle=|\psi_j(t)\rangle$ \ \ $( j=1,2,...,D )$
satisfy $i\frac{\partial}{\partial
t}|\psi_j^t\rangle=H_0|\psi_j^t\rangle,$ i.e., $|\psi_j^t\rangle$ is
a state at time $t$ evolving from $|\psi_j\rangle$ driven by $H_0.$
Clearly, $V_b\geq 0$ with equality only when $\rho\in
\mathcal{H}_{DFS}$. Taking the derivative of $V_b$ with respect to
time, we have
\begin{equation}
\dot{V}_b=-\frac 1 D\left [\sum_{n}^D\sum_m^F
f_m(t)\langle\psi_n^t|[-iH_m,\rho]|\psi_n^t\rangle+\sum_n^D\langle\psi_n^t|\mathcal
L(\rho)|\psi_n^t\rangle\right ].
\end{equation}
By the same procedure, the control fields can be established,
\begin{eqnarray}
f_{n_0}^b(t)&=&-\frac{\sum_n^D\langle\psi_n^t|\mathcal
L(\rho)|\psi_n^t\rangle}{\sum_n^D\langle\psi_n^t|[\rho,iH_{n_0}]|\psi_n^t\rangle},\nonumber\\
f_n^b(t)&=&\sum_m^D\langle\psi_m^t|[-iH_n,\rho]|\psi_m^t\rangle,
\mbox {for}  \ \ n\neq n_0. \label{conf2}
\end{eqnarray}
We observe from Eq.(\ref{nle1}) and Eq.(\ref{conf2}) that $f_n(t)$
and $f_n^b(t)$ are equivalent by setting
$\rho_D=1/D\sum_{j=1}^D|\psi_j^t\rangle\langle \psi_j^t|,$ which can
be understood as a  density matrix in the DFS.

 Discussions on this set of nonlinear equations Eq.(\ref{nle1})
are  in order as follows. By the LaSalle's invariant
principle\cite{lasalle61}, the autonomous dynamical system
Eq.(\ref{nle1}) converges to an invariant set defined by
$\mathcal{E}=\{\dot{V}=0\}$. This set is in general not empty and of
finite dimension, indicating that it is difficult to control a
quantum system from an arbitrary initial state to a given target
state. Nevertheless, by elaborately designing the control
Hamiltonian, we can control a quantum system to evolve into a
desired  subspace by  Lyapunov control. From the deviation, we find
that for our dynamical system, the invariant set $\mathcal E$ is
equivalent to $f_n(t)=0$(for any $n\neq n_0$), namely,
\begin{eqnarray}
&\ \ &\mbox{Tr}(H_n\rho\rho_D)=\mbox{Tr}(H_n\rho_D\rho), \mbox{for}
\ n\neq n_0
  \label{incon}
\end{eqnarray}
gives necessary conditions for the invariant set $\mathcal E$.  Eq.
(\ref{incon}) shows that the invariant set $\mathcal E$ depends on
both the Hamiltonian and the target state $\rho_D$. So, the
invariant set in principle can be designed by both $\rho_D$ and the
control fields  $\{H_n\}$.

\section{example}\label{exa}
\begin{figure}
\includegraphics*[width=0.7\columnwidth,
height=0.5\columnwidth]{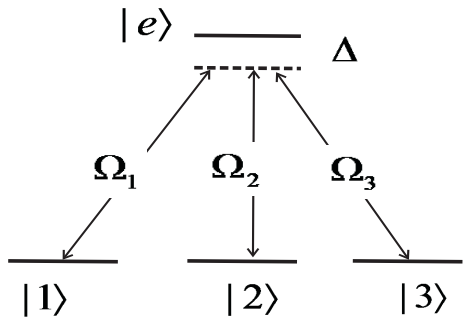}
 \caption{Atomic configurations. Three degenerate stable states
 are coupled to an excited state $|e\rangle$ by three lasers with coupling
 constants $\Omega_i$ ($i=1,2,3$). $\Delta$ denotes the detuning.}\label{f1}
\end{figure}

\begin{figure}
\subfigure{\epsfig{file=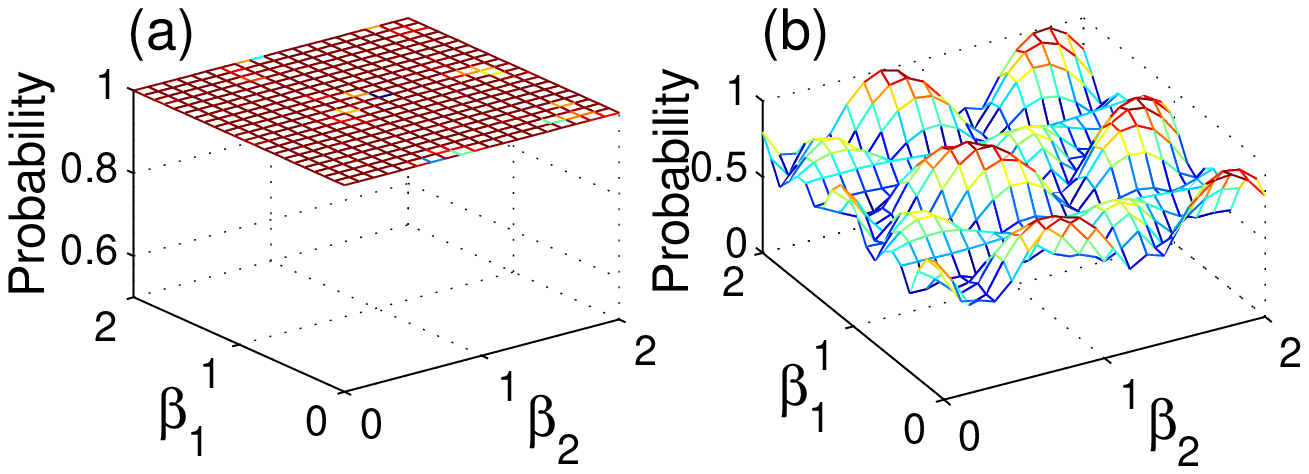,width=0.8\columnwidth,
height=0.5\columnwidth}}\vskip -2cm
\subfigure{\epsfig{file=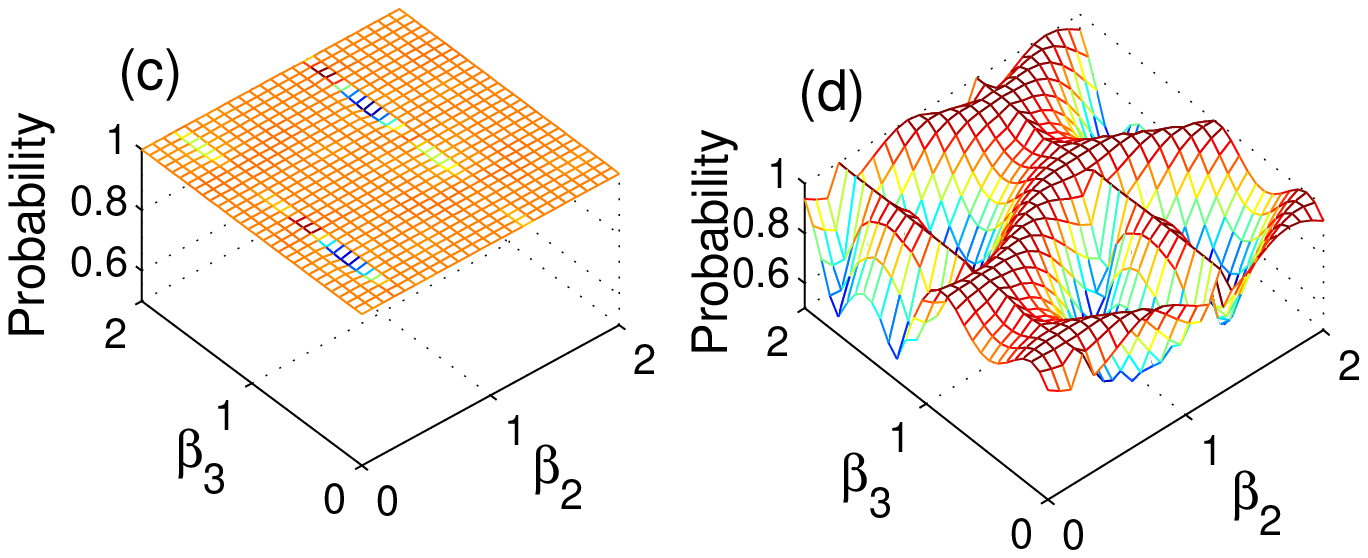,width=0.8\columnwidth,
height=0.5\columnwidth}} \vskip -1.5cm \caption{ (color online)
Probability of finding the system in DFS [(a) and (c)] and in
$|D_1\rangle$ [(b) and (d)] as a function of initial states. The
initial states are characterized by three independent parameters
$\beta_1, \beta_2$ and $\beta_3$ (in units of $\pi$). The other
parameters chosen are $\Delta=3, \Omega=5, \phi=\pi/4,$
$\theta=\pi/3,$ and $\gamma_j=0 \ \ (j=1,2,3)$. The system has
evolved for a long time enough for the system to converge. }
\label{f2}
\end{figure}

\begin{figure}
\includegraphics*[width=0.8\columnwidth,
height=0.6\columnwidth]{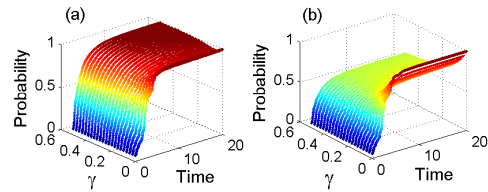}\vskip -2cm
 \caption{(color online) Probability in DFS (figure (a)), and in the dark state
 $|D_1\rangle$ {\it versus} the atomic decay and time. The initial state is
 $\beta_1=0.2\pi,$ $\beta_2=0.35\pi,$ and $\beta_3=0.2\pi.$ The other parameters
 are the same as in Fig. \ref{f2}.}\label{f3}
\end{figure}

\begin{figure}
\includegraphics*[width=0.8\columnwidth,
height=0.6\columnwidth]{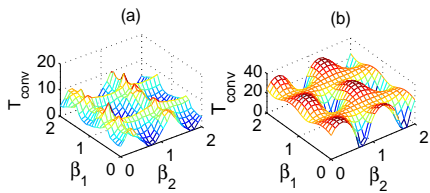}\vskip -2cm
 \caption{(color online)
 Convergence time  as a function of initial states. The
initial states are characterized by $\beta_1$, $\beta_2$ and
$\beta_3$  via Eq.(\ref{inis})  with fixed $\beta_3=0.25\pi$ in this
plot. The other parameters chosen are $\Delta=3, \Omega=5,
\phi=\pi/4,$ $\theta=\pi/3.$ (a) is drawn for $\gamma_j=0 \ \
(j=1,2,3)$ with the control, in  (b) the  atomic decay rates were
chosen to be $\gamma_j=0.1, \ \ j=1,2,3$ but without the
control.}\label{f4}
\end{figure}

As an illustration, we show in this section that the proposal works
in the setup detailed in Fig.\ref{f1},
where three degenerate stable Zeeman ground states are coupled to an
excited state through three separate external lasers. The
Hamiltonian of such a system has the form
\begin{equation}
H_0=\Delta |e\rangle\langle e|+(\sum_{j=1}^3
\Omega_j|e\rangle\langle j|+h.c.)
\end{equation}
in the rotating frame, where $\Omega_j \  \ (j=1,2,3)$ are coupling
constants. Without loss of generality, in the following the coupling
constants are parameterized as $\Omega_1=\Omega \sin\theta\cos
\phi,$ $\Omega_2=\Omega\sin\theta\sin\phi$ and
$\Omega_3=\Omega\cos\theta,$ with
$\Omega=\sqrt{\Omega_1^2+\Omega_2^2+\Omega_3^2}.$  The excited state
$|e\rangle$ is not stable, it decays to the three degenerate ground
states with rates $\gamma_1$, $\gamma_2$ and $\gamma_3$,
respectively. We assume this process is Markovian and can be
described by the Liouvillian,
\begin{equation}
\mathcal L(\rho)=\sum_{j=1}^3\gamma_j(\sigma_j^-\rho\sigma_j^+-\frac
1 2 \sigma_j^+\sigma_j^-\rho-\frac 1 2 \rho\sigma_j^+\sigma_j^-)
\end{equation}
with $\sigma_j^-=|e\rangle\langle j|$ and
$\sigma_j^+=(\sigma_j^-)^{\dagger}.$ It is not difficult to find
that the two degenerate dark states of the free Hamiltonian $H_0$
\begin{eqnarray}
|D_1\rangle&=&\cos\phi|2\rangle-\sin\phi|1\rangle,\nonumber\\
|D_2\rangle&=&\cos\theta(\cos\phi|1\rangle+\sin\phi|2\rangle)-\sin\theta|3\rangle,
\end{eqnarray}
form a DFS. Now we show how to drive the system into the DFS. For
this purpose, we choose the control Hamiltonian
$H^{\prime}=\sum_{j=1}^3 f_j(t)H_{j}$ with $H_{j}=(|e\rangle\langle
j|+|j\rangle\langle e|).$ We shall use Eq. (\ref{conf2}) to
determine the control fields $\{f_n(t)\}$, and choose
\begin{eqnarray}
|\Psi\rangle&=&\sin\beta_1\cos\beta_3|e\rangle+\cos\beta_1\cos\beta_2|1\rangle\nonumber\\
&+& \cos\beta_1\sin\beta_2|2\rangle+\sin\beta_1\sin\beta_3|3\rangle
\label{inis}
\end{eqnarray}
as  initial states, where $\beta_1$, $\beta_2$ and $\beta_3$ are
allowed to change independently. We should stress that for a
four-dimensional system, 15 independent real parameters are needed
to describe a state, making numerical simulations to exhaust all
possible initial states difficult. We  have performed extensive
numerical simulation with some initial states that can be written in
the form of  Eq. (\ref{inis}). Two types of numerical simulations
are presented.  Firstly, we numerically simulate the dynamical
system without atomic decay, i.e., $\gamma_1=\gamma_2=\gamma_3=0$,
selected results are presented in Fig. \ref{f2}.
Secondly, we simulate the effects of atomic decay on the convergence
of the dynamics, the results are plotted in Fig.\ref{f3}. From
Fig.\ref{f2} we find that all initial states in the form of
Eq.(\ref{inis}) converges to the DFS due to the Lyapunov feedback
control (see Fig.\ref{f2}-(a) and (c)), and the probability of the
system in $|D_1\rangle$  depends on the initial state
(Fig.\ref{f2}-(b) and (d)). Fig.\ref{f3} shows us that the atomic
decay does not always speed up the convergence (Fig.\ref{f3}-(a)),
and it also affects the probability of the system in the dark state
$|D_1\rangle$.
These simulations suggest that the quantum system can be driven into
the DFS, at least from the initial states Eq.(\ref{inis}). Noticing
that the open quantum system will evolve into the DFS driven by the
atomic decay alone, one may ask the following questions. How fast
does this control scheme move a state into the DFS? Is it faster
than the atomic decay? To answer these questions, we calculate the
convergence time $T_{conv}$ for the system without the Lyapunov
control but with atomic decay, the result is sown in
Fig.\ref{f4}-(b), in contrast, we compute the convergence time
$T_{conv}$ for the case when the atomic decay is zero, but with the
Lyapunov control (see Fig.\ref{f4}-(a)). By comparing these results,
we find that the Lyapunov control for this problem is indeed
effective.  To shed further insight on  the numerical simulations,
we now show that $\{|D_1\rangle, |D_2\rangle \}$  spans  LaSalle's
invariant subspace. Recall that $\rho_D=1/2(|D_1\rangle\langle
D_1|+|D_2\rangle\langle D_2|),$ $f_n(t)=0$ requires
$\sum_{i=1}^2\langle D_i|[\rho, iH_n]|D_i\rangle=0$ for any $n\neq
n_0$. Clearly, $\rho=|D_1\rangle\langle D_1|$ or
$\rho=|D_2\rangle\langle D_2|$ meets this requirement, therefore,
$\{|D_1\rangle, |D_2\rangle \}$ spans  LaSalle's invariant set.

\section{conclusion}\label{con}
In summary, we have proposed a scheme to drive an open quantum
system into the decoherence-free subspaces. This scheme works also
for closed quantum system, by replacing the DFS with a desired
subspace. This study was motivated by the fact that for a nonlinear
system, it is usually difficult to optimally control  the system
from an arbitrary initial state to a given target state. Our present
case study suggests that it is possible to drive a quantum system to
a set of states. To demonstrate the proposal we exemplify a
four-level system and numerically simulate the controlled dynamics.
The dependence of the convergence time as well as the distribution
of the system in the DFS are calculated and discussed. A comparison
between the cases with and without the Lyapunov control is also
presented. These results suggest that by designing control
Hamiltonian and target state, we can move a quantum open system into
a desired subspace.

\ \ \ \\
This work is supported by  NSF of China under grant No.
10775023, and  the National Research Foundation and Ministry of
Education, Singapore under academic research grant No. WBS:
R-710-000-008-271.


\begin{references}
\bibitem{note1} Finding a $n_0$ such that
$\mbox{Tr}\{\rho_D[iH_{n_0},\rho]\}\neq 0$ is required. As long as
$n_0$ is found, $f_{n_0}$ can be determined by
$f_{n_0}=-\frac{\mbox{Tr}[\rho_D\mathcal L
(\rho)]}{\mbox{Tr}\{\rho_D[\rho, iH_{n_0}]\}}.$ Now we show that
$f_{n_0}$ is real. By the definition of $\mathcal {L} (\rho)$,
$\mathcal {L} (\rho)$ is Hermite, then $\mbox{Tr}[\rho_D \mathcal
{L} (\rho)]$ is real. Identifying $\rho_D$ with a time-independent
density matrix, we have
$\mbox{Tr}\{\rho_D[\rho,iH_{n_0}]\}=\frac{\partial}{\partial
t}\mbox{Tr}(\rho_D\rho)$, where $\frac{\partial}{\partial
t}\rho=[\rho,iH_{n_0}]$ is assumed. Then
$\mbox{Tr}\{\rho_D[\rho,iH_{n_0}]\}$ is also real. By the same
procedure, we can show that all the control fields $f_n(t)$ are
real.
\end{references}
\end{document}